\documentclass[%
    reprint,
    superscriptaddress,
    amsmath,
    amssymb,
    aps,
    pra,
]{revtex4-1}

\usepackage{booktabs}
\usepackage{lipsum}
\usepackage{xcolor}
\usepackage{graphicx}
\usepackage{dcolumn}
\usepackage{bm}
\usepackage{siunitx}
\usepackage{xcolor}
\usepackage{dsfont}
\usepackage{tikz}
\usepackage{tikz-3dplot}
\usepackage{hyperref}
\usepackage{physics}
\usepackage{bbold}
\usepackage{float}

\graphicspath{{figures/}}

\hypersetup{colorlinks=true,citecolor=blue,linkcolor=blue,filecolor=blue,urlcolor=blue,breaklinks=true}

\newcolumntype{C}[1]{>{\centering\let\newline\\\arraybackslash\hspace{0pt}}m{#1}}

\begin{document}
\preprint{APS/123-QED}

\title{Unravelling physics beyond the standard model with classical and quantum anomaly detection}

\author{Julian Schuhmacher}
\email[Electronic address: ]{jsc@zurich.ibm.com}
\affiliation{IBM Quantum, IBM Research – Zurich, 8803 R\"uschlikon, Switzerland}
\author{Laura Boggia}
\altaffiliation[Current address: ]{LPNHE, Sorbonne Universit\'e, Universit\'e Paris Cit\'e, CNRS/IN2P3, Paris, France and IBM Research, Orsay, France}
\affiliation{IBM Quantum, IBM Research – Zurich, 8803 R\"uschlikon, Switzerland}
\affiliation{Institute for Theoretical Physics, ETH Z\"urich, 8093 Z\"urich, Switzerland}
\author{Vasilis Belis}
\affiliation{Institute for Particle Physics and Astrophysics, ETH Z\"urich, 8093 Z\"urich, Switzerland}
\author{Ema Puljak}
\affiliation{Department of Physics, Autonomous University of Barcelona, Cerdanyola del Vall\`es, Spain}
\affiliation{European Organization for Nuclear Research (CERN), Geneva, 1211, Switzerland}
\author{Michele Grossi}
\author{Maurizio Pierini}
\author{Sofia Vallecorsa}
\affiliation{European Organization for Nuclear Research (CERN), Geneva, 1211, Switzerland}
\author{Francesco Tacchino}
\author{Panagiotis Barkoutsos}
\altaffiliation[Current address: ]{PASQAL SAS, 2 av. Augustin Fresnel Palaiseau, 91120, France}
\author{Ivano Tavernelli}
\email[Electronic address: ]{ita@zurich.ibm.com}
\affiliation{IBM Quantum, IBM Research – Zurich, 8803 R\"uschlikon, Switzerland}
 
\date{\today}

\begin{abstract}

Much hope for finding new physics phenomena at microscopic scale relies on the observations obtained from High Energy Physics experiments, like the ones performed at the Large Hadron Collider (LHC).
However, current experiments do not indicate clear signs of new physics that could guide the development of additional Beyond Standard Model (BSM) theories.
Identifying signatures of new physics out of the enormous amount of data produced at the LHC falls into the class of anomaly detection and constitutes one of the greatest computational challenges.
In this article, we propose a novel strategy to perform anomaly detection in a supervised learning setting, based on the artificial creation of anomalies through a random process.
For the resulting supervised learning problem, we successfully apply classical and quantum Support Vector Classifiers (CSVC and QSVC respectively) to identify the artificial anomalies among the SM events.
Even more promising, we find that employing an SVC trained to identify the artificial anomalies, it is possible to identify realistic BSM events with high accuracy.
In parallel, we also explore the potential of quantum algorithms for improving the classification accuracy and provide plausible conditions for the best exploitation of this novel computational paradigm. 

\end{abstract}

\maketitle

\section{Introduction}
\label{sec:introduction}

Current approaches for the description of elementary particles rely on the standard model (SM) of particle physics~\cite{glashow1961partial,weinberg1967model,salam1968elementary,thooft1972regularization}.
Despite its experimental success, the SM is theoretically incomplete and new physics is yet to be explored~\cite{Langacker2017,beacham2019physics}. 
Since the discovery of the Higgs boson, the search for new physics, for example via experiments at the Large Hadron Collider (LHC), has eventually become one of the main focuses of research around high energy physics (HEP).

Experimental data obtained by the LHC experiments  can help addressing the theoretical shortcomings of the SM. 
If recorded, observations showing a significant deviation from the SM would indicate the existence of new physics.
However, state of the art LHC experiments do not yet show a clear indication of phenomena that could motivate and validate new theories. 
A key challenge in this quest is represented by the problem of storing and processing the amount of data produced by the LHC, amounting to {\cal O}($10^6$) collisions per second, each consisting of {\cal O}(1) MB.
This challenge will be amplified significantly with the start of the High Luminosity-LHC (HL-LHC) program planned for 2029~\cite{HL-LHC}, that will reach a larger luminosity at the price of a larger data flow.

\begin{figure*}[!ht]
    \centering
    \includegraphics[width=\textwidth]{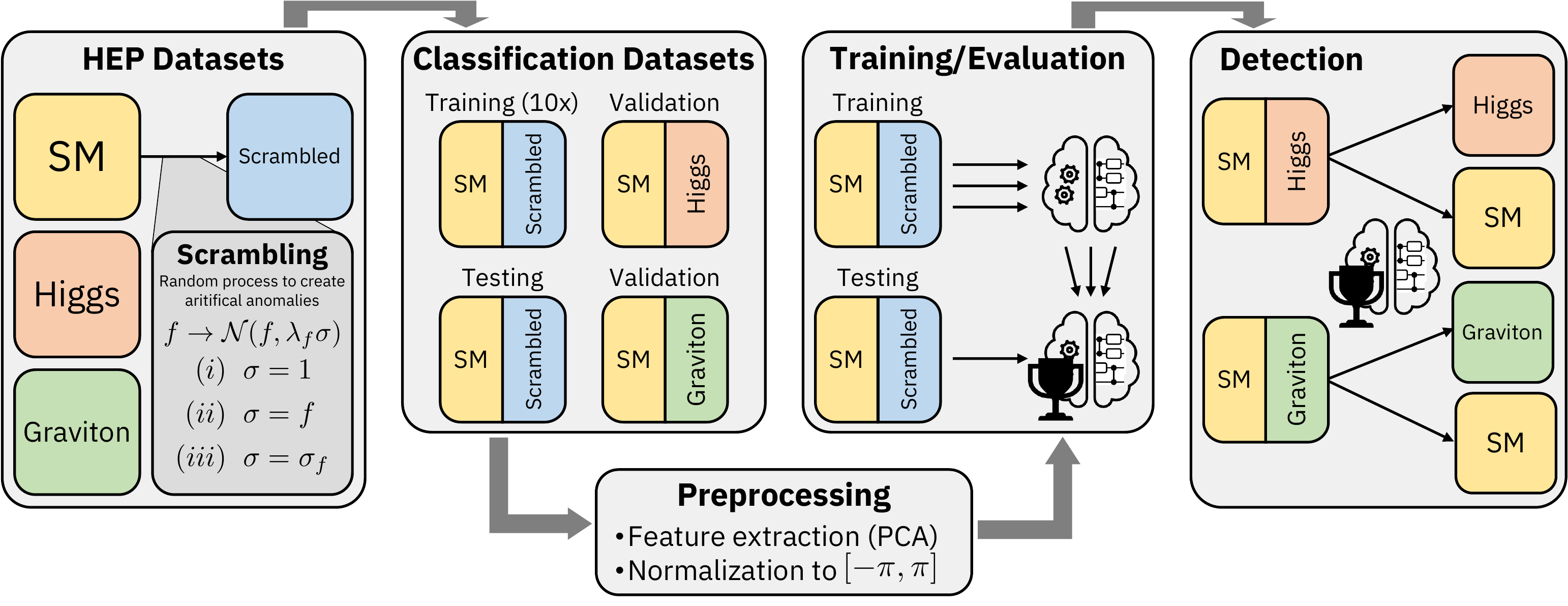}
    \caption{
        Illustration of the pipeline advised for our anomaly detection strategy. 
        We start with HEP datasets of simulated SM, Higgs and Graviton events.
        The Higgs events are considered separately, because they are not included in the processes of the SM dataset.
        We apply a random process to the SM dataset to create artificial anomalies (see~\ref{sec:data_set_scrambling}).
        Based on the four different datasets, we create balanced two-class datasets.
        One of the classes is always the SM. 
        The other class is either composed of artificial anomalies (training and testing), or Higgs or Graviton events (validation).
        We apply two preprocessing steps to the classification datasets: feature extraction with PCA and normalization of the extracted features to the interval $[-\pi,\pi]$.
        Each training dataset is then used to train multiple quantum or classical SVCs from which we select the best one based on the performance on the test dataset.
        The best SVC is then applied in the detection of unseen anomalies (Higgs or Graviton events).
    }
    \label{fig:pipeline}
\end{figure*}

A possible remedy to handle this huge amount of data relies on the use of Machine Learning (ML) models, which digest large amounts of data in order to extract underlying patterns~\cite{radovic_machine_2018,guest_deep_2018,albertsson_machine_2019,bourilkov_machine_2019,carleo_machine_2019}. 
The use of anomaly detection techniques has been proposed as a valuable tool to identify Beyond SM (BSM) events among the dominating number of SM background events~\cite{Nachman_2020,Kasieczka_2021,Aarrestad:2021oeb}.
Different approaches have been investigated. In the supervised learning setting, the ML model is trained to distinguish between background and signal events using a dataset where the events were labelled with the corresponding class. However, to acquire the labels of an event some preliminary knowledge about the studied processes is required, e.g. through numerical simulations of a BSM theory. This typically limits the generalization power of a supervised algorithm. In the context of the search for new physics at the LHC, this means that this approach is typically effective whenever the considered signal is the correct one (e.g., for Higgs boson searches).
In unsupervised learning setting, the ML model is trained to learn the structure of a dataset largely dominated by SM events. Without being provided any additional information, it aims at identifying an anomaly as an outlier of some typicality measure, learned during the training process. An unsupervised learning algorithm requires no labels. On one hand, this increases the generalization power of the model. On the other hand, it reduces the accuracy since less information about any specific signal is used. 

In this paper, we propose a strategy for anomaly detection that tries to retain the best out of the supervised and unsupervised setting. We propose a supervised learning setting in which the signal sample is built perturbing the background sample, without relying on any specific BSM theory. The background events are represented by a selection of SM processes, and signal events are generated artificially through a random process, which is denoted here as \textit{scrambling}.
This approach ensures that we introduce as little physically-inspired bias as possible into the types of signal events we are looking for, while guaranteeing that the defined processes are compliant with the conservation laws of physics and detector-specific constraints. 
We solve the resulting binary classification between background and signal events with the support vector classifier (SVC) approach.
An illustration of the proposed anomaly detection pipeline is shown in Fig.~\ref{fig:pipeline}.

In parallel with classical SVC techniques, we also consider the application of quantum kernel based classification methods.
Quantum Machine Learning (QML) has recently been proposed as a new framework offering potential speedups and performance improvements over classical ML~\cite{Biamonte2017,Guan_2021,Mangini_2021,delgado2022quantum}.
Several QML algorithms, like Quantum Support Vector Classifiers (QSVCs)~\cite{Rebentrost2014,Havlicek2019,Schuld2018}, Variational Quantum Classifiers (VQCs)~\cite{Havlicek2019,Schuld2018}, Quantum Convolutional Neural Networks (QCNNs)~\cite{cong_quantum_2019}, or quantum autoencoders~\cite{Romero_2017} have been applied to a wide range of HEP problems~\cite{WuQSVC2021,heredge2021quantum,TerashiVQC2021,gianelle2022,BlanceVQC2021,WuVQC2021,BelisVQC2021,chen2022quantum,chen2021hybrid,Ngairangbam_2022}.
With the current methods, quantum algorithms generally achieve a performance similar to their classical counterparts. 
While future quantum hardware and algorithmic improvements may lead to more significant advantages, a solid and rigorous benchmark of the available quantum techniques is already important today.

In this work, we verify numerically the feasibility to distinguish SM events from artificially created anomalies.
We show that such a classifier preserves its discrimination power once the scrambling anomalies are replaced by events from realistic BSM theories.
Furthermore, we demonstrate the application of a kernel-based quantum classification algorithm to the problem under study. 

This paper is organized as follows: 
In section~\ref{sec:data_set_scrambling} we present the scrambling method, followed by a short description of the applied classification algorithm in section~\ref{sec:svc}. 
In section~\ref{sec:results} we demonstrate and discuss the effectiveness of the proposed anomaly detection workflow.
Finally, in section~\ref{sec:conclusions}, we conclude with a general discussion of the proposed method and the application of quantum algorithms for the studied classification task.

\section{Dataset scrambling}
\label{sec:data_set_scrambling}

\begin{figure}
    \centering
    \includegraphics[width=\columnwidth]{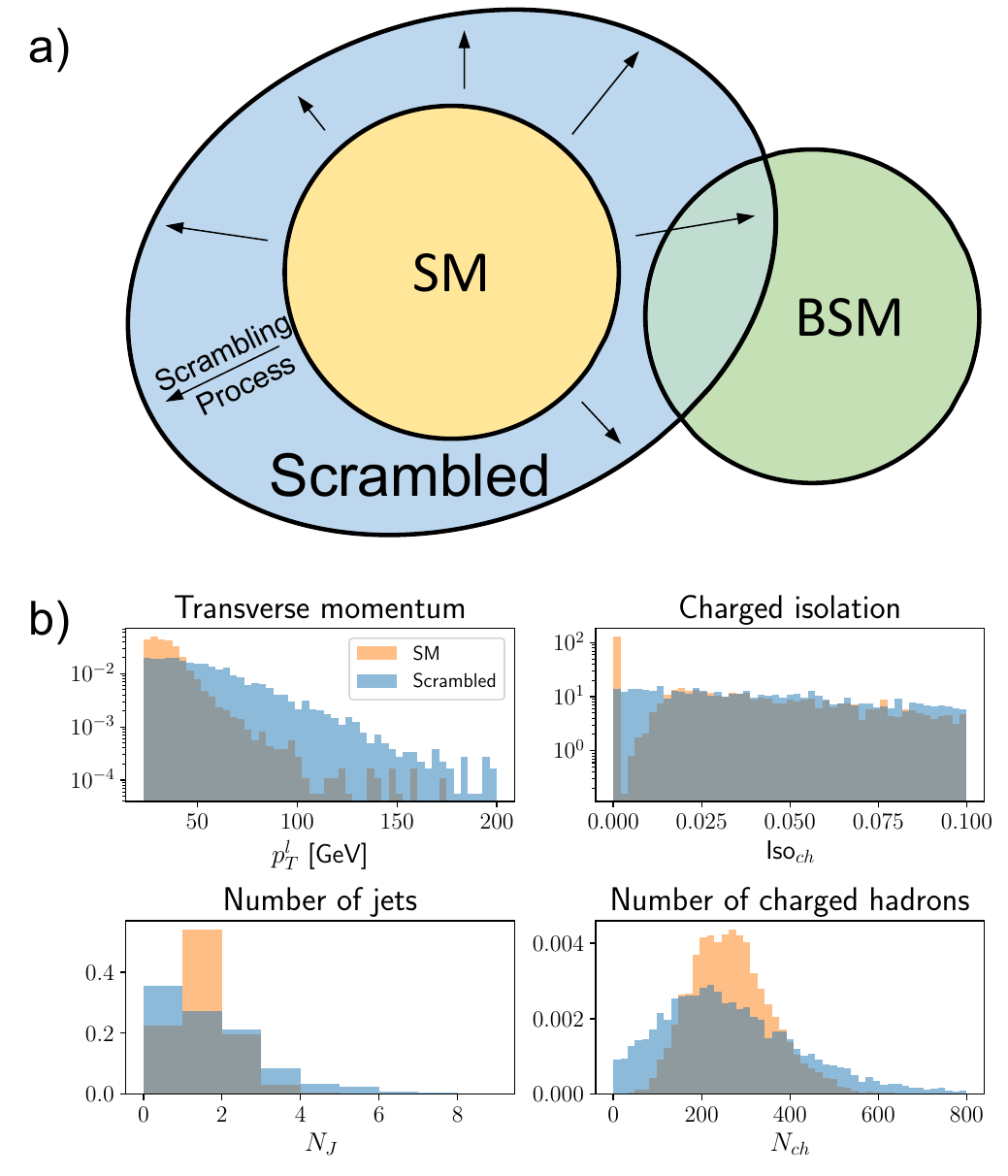}
    \caption{
        \textbf{a} Schematic illustration of the scrambling idea. 
        The scrambling process generates random events (blue region) based on SM events (yellow region). 
        Being able to distinguish a SM event from a scrambled event, enables us to identify events originating from physics beyond the SM (green region).
        \textbf{b} Comparison of initial SM data and scrambled data at medium scrambling strength, for a selection of the high-level features. 
    }
    \label{fig:scrambling}
\end{figure}

To generate a dataset for our supervised learning problem, we start from an existing collection of SM events and we vary the different features in order to introduce anomalies (artificial events). 
The events are represented by features like the momenta or the number of particles, extracted from the collision data obtained from experiments or simulations (see section~\ref{ssec:hep_datasets} and appendix~\ref{app:datasets} for details).
The variation of these features is done under certain constrains imposed by physics conservation laws or experiment-related constraints. 
We call this process data scrambling. 
Its goal is to generate events that do not conform with the SM, without relying on a specific BSM theory, and therefore introducing as little bias as possible about the type of BSM events we would like to identify.
We propose to do this via a random perturbation of the SM events.
The main idea of the scrambling is illustrated in Fig.~\ref{fig:scrambling}a.
Starting from the SM dataset (yellow region), we generate artificial events outside the SM (blue region). Even though there is an overlap between the region of SM events and artificial events, we want to verify that a classifier, trained to separate the SM events from the artificial BSM ones, would retain its discrimination power once applied to realistic BSM events (green region).

The scrambling is done by replacing a feature in the original SM dataset with a new value chosen according to a Gaussian distribution $\mathcal{N}(\mu, \sigma)$, where $\mu = f$ is the initial value of the feature and $\sigma$ is the standard deviation. 
Depending on the feature, the standard deviation of the scrambling distribution is chosen according to one of the following three options;
\textit{(i)} the standard deviation is fixed to a constant, $\sigma = \lambda_f$, \textit{(ii)} proportional to the standard deviation $\sigma_f$ of the initial feature distribution, $\sigma = \lambda_f \sigma_f$, or \textit{(iii)} proportional to the feature value $f$, $\sigma = \lambda_f f$.
The constant $\lambda_f$, in the following denoted as \textit{scrambling factor}, determines the strength of the scrambling. \\
Some of the features are correlated (e.g. transverse momentum of lepton and missing transverse energy) and therefore cannot be scrambled individually.
Additionally, depending on the feature we have to implement different strategies to respect conservation laws or detector-specific limitations. 
Therefore, the features are divided into four categories: momenta, isolations, jets and particle numbers, each with their own scrambling strategy.
The scrambling strategies are presented in detail in appendix~\ref{app:scrambling}.
In Fig.~\ref{fig:scrambling}b, the scrambling is visualized for some features in the SM dataset, each belonging to one of the categories mentioned above.

The idea of the scrambling is not limited to the specific choice of the sampling distributions introduced above. 
In principle, any sampling distribution is valid, as long as some generated events lie outside the ``event space'' of the SM, and they respect the physical conservation laws and the constraints imposed by the detector. 
However, there is no guarantee that any chosen scrambling distribution will generate events resembling BSM events.
Nevertheless, the hope is that by learning to distinguish between SM and artificial events, we obtain some level of generalization on out-of-distribution samples, and therefore the possibility to detect BSM events even if they would lie outside the space of scrambled events (no overlap between the blue and green region in figure~\ref{fig:scrambling}a).

\section{Support Vector Classifier}
\label{sec:svc}

\begin{figure}
    \centering
    \includegraphics[width=0.8\columnwidth]{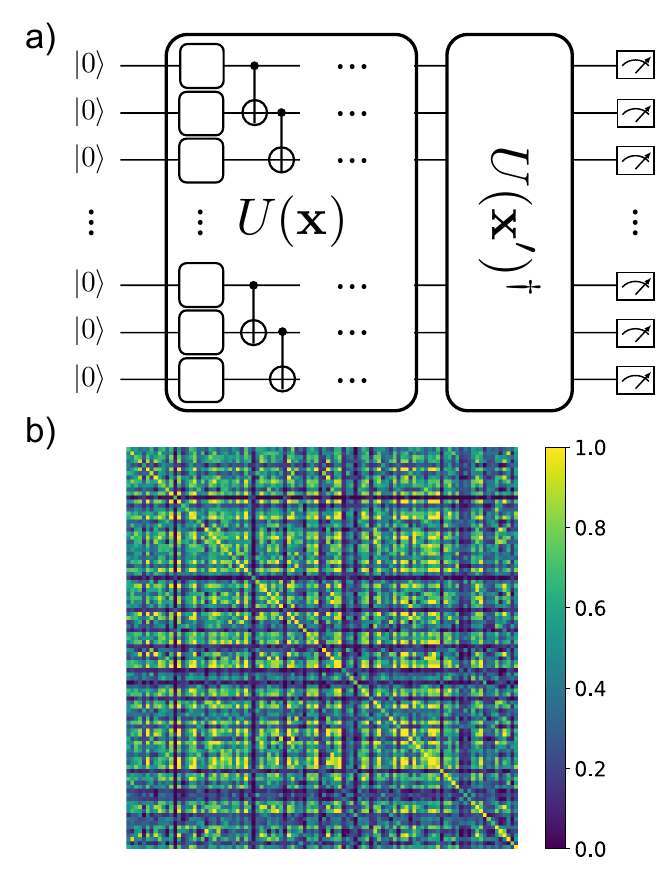}
    \caption{
        \textbf{a} Quantum circuit used by the Quantum Support Vector Classifier (QSVC) to measure the overlap between two encoded quantum states. 
        The feature vector $\bm{x}$ is encoded into a quantum state through the parameterized unitary $U(\bm{x})$.
        \textbf{b} Kernel matrix for the classification between SM events and artificial anomalies calculated on the IBM quantum processor \textit{ibm\_cairo} using 6 features (corresponding to 6 qubits).
    }
    \label{fig:quantum_classifier}
\end{figure}

\paragraph{Classical SVC}
The Support Vector Classifier (SVC), which belongs to the family of kernel methods, is a supervised learning model that can draw hyperplanes between two classes of data points.
By embedding data points into a high dimensional feature space, where they become linearly separable, SVCs can successfully solve complex classification tasks.
The success of the SVC results from the so-called \textit{kernel trick}, which allows to calculate a similarity measure between data points (i.e., the kernel) without explicitly performing the mapping to the high dimensional feature space.
An example of such a kernel is the widely used radial basis function (RBF) kernel (also known as Gaussian kernel),
\begin{equation}\label{eq:classical_kernel}
    K(\mathbf{x}_i, \mathbf {x}_j) = \exp \left(-\gamma \|\mathbf{x}_i - \mathbf{x}_j \|^{2} \right) \,,
\end{equation}
where $\mathbf{x}_i,\mathbf{x}_j \in \mathbb{R}^n$, and $\gamma$ is a hyperparameter to determine the bandwidth of the kernel function. 
Notice that this specific kernel function corresponds to a situation in which data points would effectively be mapped into an infinite-dimensional feature space~\cite{Amnon_2009}.

\paragraph{Quantum SVC}
\label{sssec:qsvc}
For the quantum SVC (QSVC), the kernel values are evaluated in a Hilbert space of quantum states. 
Specifically, classical features $\mathbf{x} \in \mathcal{X}$ are encoded in the quantum state space $\mathcal{F}$ via a feature map $\phi: \mathcal{X} \rightarrow \mathcal{F}$,
\begin{equation}
    \mathbf{x} \rightarrow \rho(\mathbf{x}) = \ketbra{\phi(\mathbf{x})} \,.
\end{equation}
Usually, the feature map is given in terms of a parameterized unitary $U(\mathbf{x})$ applied to a fixed reference state, e.g. $\ket{\phi(\mathbf{x})} = U(\mathbf{x}) \ket{0}$.
The kernel used in the classical SVC optimization is then calculated as the overlap between two encoded quantum states,
\begin{equation}\label{eq:qsvc}
    K(\mathbf{x}_i, \mathbf{x}_j) = \Tr\left[ \rho(\mathbf{x}_i) \rho(\mathbf{x}_j) \right] = \left| \braket{\phi(\mathbf{x}_i)}{\phi(\mathbf{x}_j)} \right|^2 \,.
\end{equation}
A kernel of this form can be evaluated on a quantum device, and could bring an advantage over classical SVCs provided that the quantum feature map is hard to simulate classically~\cite{Havlicek2019,Liu2021rigorous}.
The quantum circuit used to evaluate the kernel is schematically shown in figure~\ref{fig:quantum_classifier}a.
The overlap between the encoded quantum states is given by the probability of measuring the all-zero state at the end of the circuit.
The details about the applied feature map are presented in appendix~\ref{app:feature_map}.
An example of a kernel matrix resulting from such a quantum feature map is shown in figure~\ref{fig:quantum_classifier}b.
The kernel values were estimated with the IBM Quantum processor \textit{ibm\_cairo} using 6 input features (corresponding to 6 qubits).

\section{Results} 
\label{sec:results}

\subsection{Datasets}
\label{ssec:hep_datasets}

In the studied anomaly detection problem, background events are represented by simulated samples of the SM processes typically observed at 13 TeV~\cite{cerri_variational_2019,Knapp2021}.
The included processes are (with relative occurrences): W bosons decaying into a charged and a neutral lepton (59.2\%), multi-jet production from QCD processes (33.8\%), Z boson decaying into two charged leptons (6.7\%), and $t\bar{t}$ production (0.3\%). 
The processes are described by 23 high-level features representing simulated measurement results as obtained with e.g. the CMS detector.
A detail list of the features and their description in given in appendix~\ref{app:datasets}.
For the validation of the proposed anomaly detection strategy, we use two different types of processes as signal events. 
The first one is the Higgs boson, which has not been included in the SM dataset, and can therefore be interpreted as a BSM particle. 
The dataset for the Higgs boson consists of simulated high-mass Higgs particle produced via vector boson fusion~\cite{dataset_higgs}. 
Additionally, we also use a sample of simulated Randall-Sundrum Gravitons~\cite{Randall:1999ee} decaying to two Z bosons, forcing each Z to decay to a lepton pair~\cite{dataset_graviton}. 
In both BSM datasets, the events are represented by the same 23 features as in the SM dataset.

Using the scrambling process introduced in section~\ref{sec:data_set_scrambling} we create three different datasets with artificial anomalies, each with different scrambling intensity, denoted as \textit{low}, \textit{medium} and \textit{high} scrambling.
The corresponding scrambling factors $\lambda_f$ are listed in appendix~\ref{sapp:scrambling_factors}. 
Here, we limit ourselves to the scrambling of a subset of 17 out of the 23 high level features, allowing us to satisfy physical constraints like energy conservation.

Following this methodology, we create training, testing and validation datasets for the binary classification. 
Unless stated otherwise, the classification datasets contain 1000 samples per class, where the background is labelled as the negative class, and the signals as the positive class. 
We create 10 training datasets for each scrambling strength, and one test dataset, all consisting of a combination between SM data and artificial anomalies. 
Further, we prepare two validation datasets, one consisting of a combination of SM and Higgs data, and the other of a combination of SM and Graviton data.
The datasets and their composition are schematically shown in the first two panels of Fig.~\ref{fig:pipeline}.

\subsection{Numerical experiments}

The training of a classifier consists of two main steps, the preprocessing of the data samples and the training itself. 
For the preprocessing we consider the following steps: standardization applied of the input features, feature selection or feature extraction to reduce the number of features used in the classification, and normalization applied to balance the importance between the feature before forwarding them to the classifier.
In the classical case, the kernel matrix for the training of the classifier is calculated with equation~\eqref{eq:classical_kernel}.
To calculate the kernel matrix with a quantum computer, we use a parameterized quantum circuit to encode the classical features in a quantum state, and obtain the kernel values by calculating the fidelity between two encoded data samples (equation~\eqref{eq:qsvc}).
Similarly to the classical case, and as proposed in~\cite{shaydulin_importance_2021,peters_machine_2021}, we introduce a hyperparameter $\gamma$ in the quantum feature map to control its resolution in the Hilbert space. 
Figure~\ref{fig:training_workflow} in the appendix shows an overview of the training workflow, and in appendix~\ref{app:training_workflow} we present a detailed hyperparameter optimization for the QSVC model.

For the simulations we have fixed the steps of the training workflow in the following way.
We use no standardization transformation prior to the feature extraction with PCA, since the tested standardization algorithms all lead to a reduced performance (see appendix~\ref{sapp:standardization}). 
In the normalization step we scale all features to the interval $[-\pi,\pi]$.
For the quantum classifier, we encode the data with a feature map similar to the one introduced in Ref.~\cite{Havlicek2019} with a hyperparameter fixed to $\gamma = 0.5$.
A detailed description of the applied feature map and a justification for fixing its hyperparameter are given in appendix~\ref{app:feature_map} and~\ref{sapp:scaling_factor}, respectively.
In the case of the classical SVC, the hyperparameter of the radial-basis function kernel (equation~\eqref{eq:classical_kernel}) is optimized for each classification individually, by selecting the one that achieves the best validation score on an independent test dataset.

\begin{figure*}
    \centering
    \includegraphics[width=\textwidth]{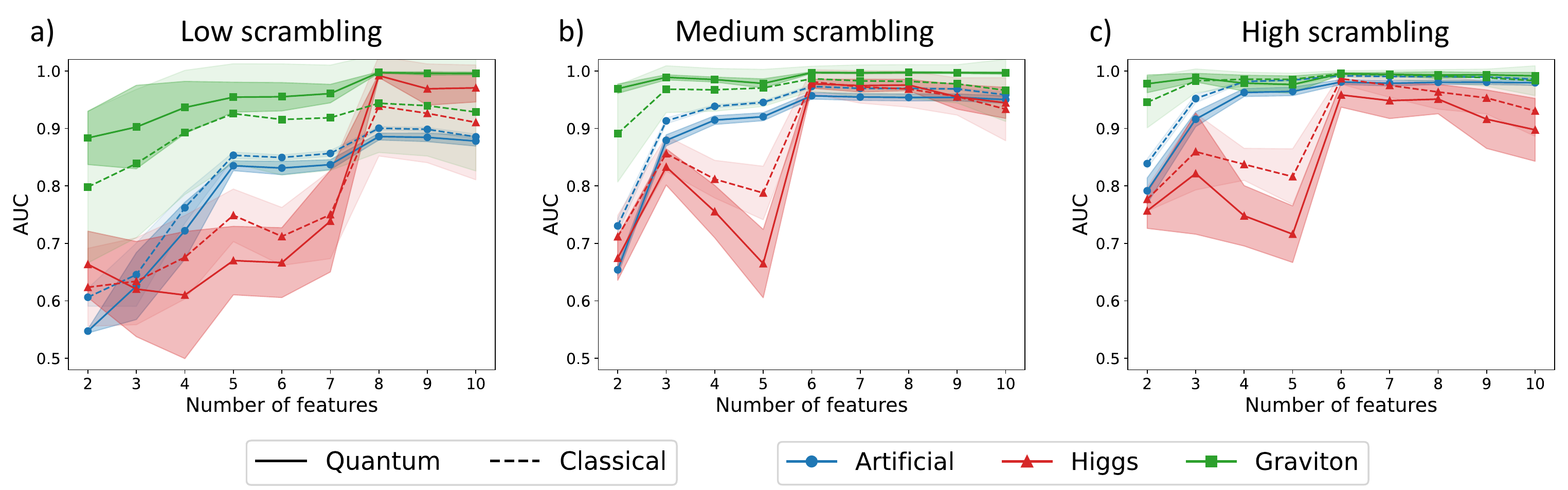}
    \caption{Validation AUC for classification of BSM events among SM events with quantum and classical SVCs for different scrambling strength and different number of features.
    \textbf{a} Validation AUC for identifying artificial anomaly (blue), Higgs (red) or Graviton (green) events among SM events for a QSVC (solid lines) and a classical SVC (dashed lines). The classifiers were trained on artificial anomalies generated with the low scrambling strength.
    \textbf{b, c} Equivalent results, but for medium and high scrambling intensity, respectively.}
    \label{fig:results}
\end{figure*}

Using this workflow, we train (Q)SVCs for each scrambling strength and training dataset, and evaluate them on the corresponding test and validation datasets.
All quantum computations were done with Qiskit~\cite{Qiskit}.
The results are shown in figure~\ref{fig:results}.

\begin{table*}
    \centering
    \begin{tabular}{ c | c | c | c | c | c }
         & Scrambling & Number of & \textbf{Anomalies} & \textbf{Higgs} & \textbf{Graviton} \\
        Algorithm & strength & features & AUC & AUC & AUC \\
        \hline
                      &    low & 8 & 0.900 $\pm$ 0.003 & 0.939 $\pm$ 0.087 & 0.944 $\pm$ 0.086 \\
         \textbf{SVC} & medium & 6 & 0.971 $\pm$ 0.003 & 0.981 $\pm$ 0.021 & 0.986 $\pm$ 0.022 \\
                      &   high & 6 & 0.992 $\pm$ 0.001 & 0.987 $\pm$ 0.010 & 0.995 $\pm$ 0.006 \\
        \hline
                      &    low & 8 & 0.886 $\pm$ 0.006 & 0.992 $\pm$ 0.003 & 0.997 $\pm$ 0.001 \\
        \textbf{QSVC} & medium & 6 & 0.957 $\pm$ 0.006 & 0.978 $\pm$ 0.008 & 0.997 $\pm$ 0.001 \\
                      &   high & 6 & 0.981 $\pm$ 0.004 & 0.958 $\pm$ 0.021 & 0.996 $\pm$ 0.002 \\
        \hline
        \textbf{QSVC}* & medium& 6 & 0.870 $\pm$ 0.052 \\
    \end{tabular}
    \caption{Highest validation AUC for classification between artificial anomalies and SM events with quantum and classical SVCs for different scrambling strengths (fourth column). 
    The third column lists the number of features for which this validation AUC is achieved.
    The last two columns hold the corresponding detection AUC of the Higgs and Graviton events. \\
    *Quantum kernel estimation executed on IBM Quantum device \textit{ibm\_cairo}.}
    \label{tab:validation_auc}
\end{table*}

In figure~\ref{fig:results}, we compare the AUC score for identifying artificial anomalies (blue lines), Higgs events (red lines) and Graviton (green lines) events with a quantum (solid lines) and a classical (dashed lines) classifier at different scrambling strength and for different number of features. \\
A few observations are in order: 
confirming our expectations, the validation score for identifying the artificial anomalies (blue lines) increases with the number of features, and the scrambling strength.
We can also confidently conclude that the proposed anomaly detection strategy is successful, 
since it generalizes to Higgs and Graviton events, even though the classifiers were trained to identify the artificial anomalies.
For the low scrambling strength, using 8 features leads to the highest identification AUC of artificial anomalies and the highest detection AUC of the Higgs and Graviton events.
For the medium and high scrambling strengths the highest AUC values are reached for 6 features. 
A possible reason why the number of features with the highest AUC changes for the different scrambling factors is that inducing more extreme anomalies produces, on average, events which are easier to distinguish from SM ones.
In general, this leads to both an increased validation and detection AUC for the same number of features.
In other words, for an increased scrambling strength, a lower number of features is required to achieve the same performance. 
However, the highest detection AUC value is reached for the low scrambling.

In most cases the classifiers are even better at detecting the Higgs and the Graviton events than identifying the artificial ones.
The numerical values of the highest validation AUC and the corresponding detection AUC are listed in table~\ref{tab:validation_auc}.

The performance of the classical and quantum SVCs is very similar.
Looking only at the classification between SM and artificial anomalies (blue curves), the classical SVC outperforms the quantum SVC in all cases. 
However, the gap gets smaller for increasing number of features, and increasing scrambling strength.
Focusing on the detection we observe the contrary behaviour.
For low scrambling strength (where the detection AUC is the highest) the quantum SVC is better at detecting the Higgs and Graviton events. 
However, increasing the scrambling strength closes the gap between the classical and quantum SVC when detecting Graviton events, and the order gets reversed for the detection of Higgs events.

Overall, the results suggest that, although a quantum SVC can be better than a classical one in terms of detection ability, in general the two methods exhibit essentially comparable performances.

A possible explanation for the improved detection score of the QSVC for Higgs and Graviton events -- which are not explicitly present in the dataset of anomalies -- could lie in lower overfitting on the classification task and, hence, a better generalization power.
However, we have observed that introducing a bias to the classical SVC by fixing the hyperparameter of the classical kernel to $\gamma = 0.5$, only leads to a minimal drop in its classification accuracy, but to a significant gain in the detection accuracy.
The corresponding figures are shown in appendix~\ref{sapp:classical_bias}.

\subsection{Hardware experiments}
\label{ssec:hardware_experiments}

\begin{figure}
    \centering
    \includegraphics[width=0.8\columnwidth]{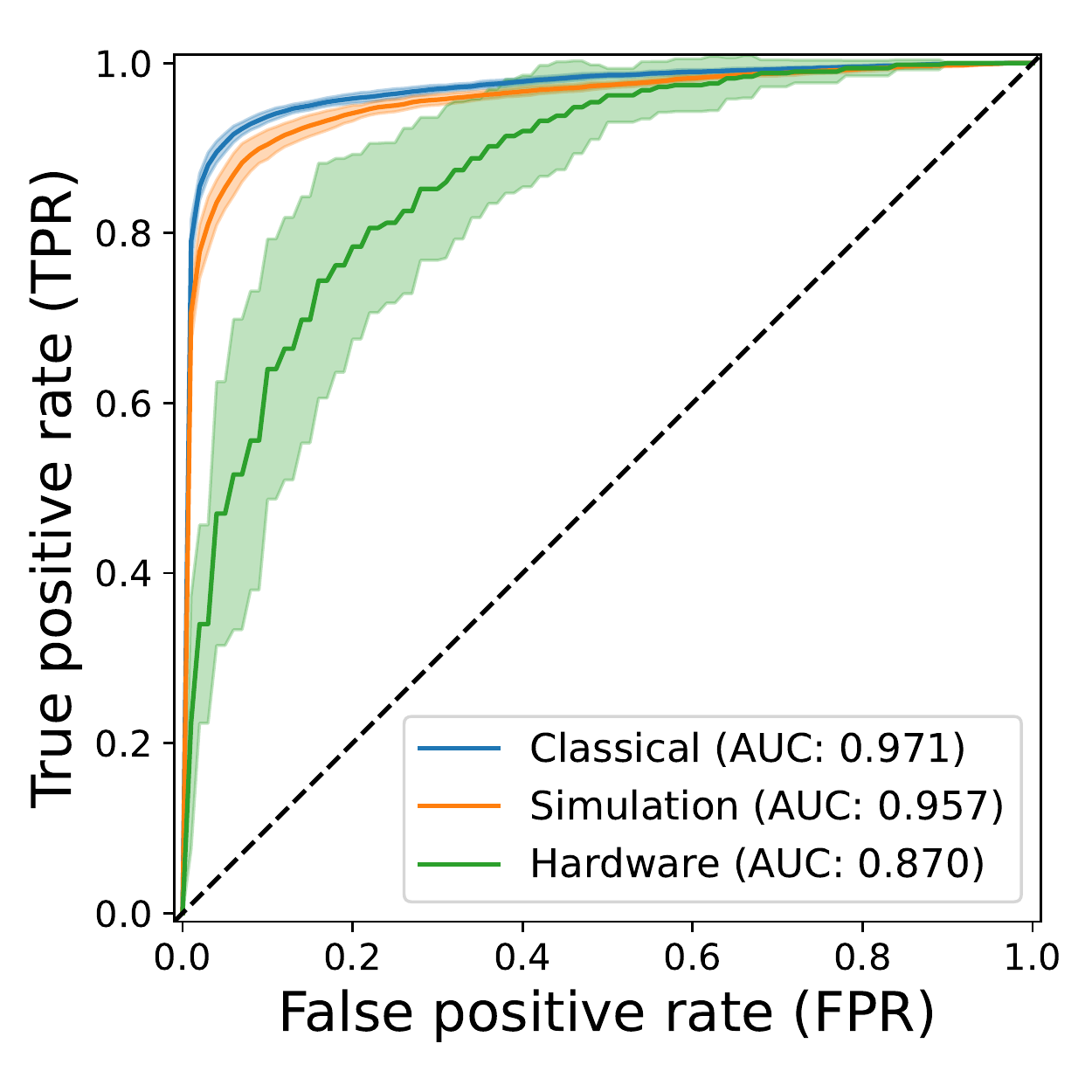}
    \caption{
        ROC-AUC curve of the classification between SM events and artificial anomalies. 
        The kernel matrices for the classification were provided by a classical kernel function (blue), a simulated quantum kernel (orange), and a quantum kernel estimated using the quantum device \textit{ibm\_cairo} (green).
    }
    \label{fig:roc_hardware}
\end{figure}

For all the results presented in the previous sections, the quantum kernel values were computed via the simulation of a perfect quantum computer, without errors due to finite measurement statistics or hardware noise. 
While the former can be included in numerical simulations without much effort~\cite{schuhmacher2022}, the latter is harder to capture, and it is therefore important to benchmark quantum algorithms directly on existing quantum processors.

We perform all hardware calculations on the IBM Quantum superconducting device \textit{ibm\_cairo}, using the same experimental setup as above (same training workflow and hyperparameters).
We use $10^4$ repetitions (shots) for the estimation of the kernel values, and we apply a depolarization error mitigation method to the obtained kernel matrices (see appendix~\ref{sapp:depol_mitigation}).
To ease the requirements on the quantum device, we only use 50 events per class and only look at the classification between SM and artificial events.
Additionally, we also just consider the 6-feature case for medium scrambling strength.

One instance of a kernel matrix calculated with the quantum device is displayed in figure~\ref{fig:quantum_classifier}b.
In figure~\ref{fig:roc_hardware}, we also report the average ROC-AUC curve of the quantum SVC trained on noisy kernel matrices. 
Performances are compared with the simulated quantum SVC and the classical SVC.
The validation AUC of the QSVC is listed in table~\ref{tab:validation_auc}.

The validation AUC evaluated on the hardware is lower than the AUC obtained with classical SVC and simulated quantum SVC.
The most probable reasons for this discrepancy are hardware noise sources other than the mitigated depolarization errors, and the lower number of training and testing samples used in the hardware experiments.

\section{Conclusions}
\label{sec:conclusions}

In this work, we proposed and successfully tested a supervised learning strategy for anomaly detection. 
Instead of generating anomalies provided by a BSM theory we randomly generate them with a scrambling process that is based on an initial dataset containing SM processes. 
We verify the feasibility of the proposed anomaly detection strategy by identifying HEP processes that were not included in the SM dataset during the training of a classifier.
The success of the strategy can be confirmed with both classical and quantum SVCs approaches.

Currently, the scrambling process generates events which have a high overlap with the initial data samples, and reducing this overlap could improve the classification between SM and artificial events.
However, this does not guarantee that the detection of unknown events will also improve.
A possible strategy to reduce the overlap would be to use a scrambling distribution that is different from the Gaussian distribution.
The only constraints that the generated events have to respect are physical conservation laws and requirements by the detection.
Therefore, it would in principle be possible to construct a scrambling process that generates events with less overlap to the initial data samples, or even generate events that lie in a desired region of the ``event space''.
Additionally, we would like to stress that the proposed scrambling strategy is not limited to HEP datasets, and could in principle be applied for any anomaly detection problem.

While our results establish empirically a successful application of quantum kernel methods to a HEP anomaly detection task, we could not yet observe a generalized promise for quantum advantage. 
However, it is not possible to rule out individual problem instances where a quantum classifier could outperform a classical one, e.g. for different scrambling distributions, or a higher number of features.
In fact, a similar study on anomaly detection in HEP has found evidence for scenarios where a quantum model can outperform the classical counterpart~\cite{Wozniak_2023}.
Such results emerge from a combination of unsupervised learning approaches applied to a dataset where the collision events are described by the 4-momenta of the involved particles, hence using a representation closer to the physical ``raw data`` compared to the dataset used in this work.
Generally, there is mounting evidence that quantum advantage on classical datasets can only be strictly guaranteed when specific structure is present~\cite{Liu2021rigorous,aaronson2022,kubler2021inductive,slattery2022numerical}. An analysis of our classification problem via the methods proposed in Ref.~\onlinecite{Huang2021} is also not conclusive (see appendix~\ref{sapp:potential_qa}).

As the data constituting the target of our work originates from a quantum HEP process, the idea of using a quantum technique for its classification seems rather natural.
However, the features currently used to describe the processes, collected from detector measurements, are fully classical. 
This loss of ``quantumness'' could in fact represent an important limitation to the use of more sophisticated QML techniques for the analysis and classification of quantum states~\cite{cong_quantum_2019,Tacchino2020IEEE,wu2021,huangQA2022}. 
We therefore believe that a different setup bypassing the extraction of classical features could be a promising road towards quantum advantage also in the context of HEP.
While for collision events at LHC such a setting is currently not possible, there already exist experiments at CERN where quantum sensors are studied for information extraction~\cite{colaleo2021}.
In the future, it could certainly be interesting to couple a quantum processor to quantum sensors embedded in detectors, hence enabling direct manipulation and classification of quantum amplitudes produced in an experiment.

\section{Acknowledgements}
J.S. and I.T. acknowledge financial support from the Swiss National Science Foundation (SNF) through the grant No. 200021-179312. V.B. is supported by an ETH Research Grant [grant No. ETH C-04 21-2]. M.G. and S.V. are supported by the CERN Quantum Technology Initiative. M.P. and E.P. are supported by the European Research Council (ERC) under the European Union's Horizon 2020 research and innovation program (grant agreement n$^o$ 772369).
We acknowledge the use of IBM Quantum services for this work. IBM, the IBM logo, and ibm.com are trademarks of International Business Machines Corp., registered in many jurisdictions worldwide. Other product and service names might be trademarks of IBM or other companies. The current list of IBM trademarks is available at \url{https://www.ibm.com/legal/copytrade}.

\bibliography{references/hep,references/hep_ml,references/ml,references/qml,references/hep_qml}
\bibliographystyle{unsrt}

\appendix

\section{HEP Datasets}
\label{app:datasets}

The events in the HEP datasets are described by 23 high-level features~\cite{cerri_variational_2019,Knapp2021}. 
The features are listed in table~\ref{tab:features}, including a column which indicates if the feature is considered in the scrambling process.

\begin{table}[h!]
    \centering
    \begin{tabular}{ l | p{0.6\columnwidth} | c }
        Feature & Description & Scrambled? \\
        \hline
        $H_T$ & The scalar sum of the transverse momenta $p_T$ of all jets having $p_T > 30$~GeV and $|\eta| < 2.4$. & x \\
        $M_J$ & The invariant mass of all jets entering the $H_T$ sum. & \\
        $N_J$ & The number of jets entering the $H_T$ sum. & x \\
        $N_B$ & The number of jets identified as originating from a $b$ quark. & x \\
        $p^\mu_{T,TOT}$ & The vector sum of the $\bm{p}_T$ of all PF muons in the event having $p_T > 0.5$~GeV. & x \\
        $M_\mu$ & The combined invariant mass of all muons entering the sum in $p^\mu_{T,TOT}$. \\
        $N_\mu$ & The number of muons entering the sum in $p^\mu_{T,TOT}$. & x \\
        $p^e_{T,TOT}$ & The vector sum of the $\bm{p}_T$ of all PF electrons in the event having $p_T > 0.5$~GeV. & x \\
        $M_e$ & The combined invariant mass of all electrons entering the sum in $p^e_{T,TOT}$. \\
        $N_e$ & The number of electrons entering the sum in $p^e_{T,TOT}$. & x \\
        $N_{neu}$ & The number of all neutral hadron PF-candidates. & x \\
        $N_{ch}$ & The number of all charged hadron PF-candidates. & x \\
        $N_\gamma$ & The number of all photon PF-candidates. & x \\
        $p^l_T$ & The transverse momentum of the highest $p_T$ lepton in the event. & x \\
        $\eta_l$ & The lepton pseudorapidity. & x \\
        $q_l$ & The lepton charge (either $-1$ or $+1$). & \\
        $\text{Iso}^l_{ch}$ & The lepton isolation related to all other charged hadron PF-candidates. & x \\
        $\text{Iso}^l_{neu}$ & The lepton isolation related to all neutral hadron PF-candidates. & x \\
        $\text{Iso}^l_{\gamma}$ & The lepton isolation related to all photons. & x \\
        $\text{MET}_\parallel$ & The parallel component of the missing transverse energy with respect to the lepton. & x \\
        $\text{MET}_\bot$ & The orthogonal component of the missing transverse energy with respect to the lepton. & x \\
        $M_T$ & The combined transverse mass of the lepton and the missing transverse energy system. & \\
        $\text{IsEle}$ & A flat set to 1 if the lepton is an electron, 0 if it is a muon. & \\
    \end{tabular}
    \caption{High-level features used as description of the events in the HEP datasets~\cite{cerri_variational_2019,Knapp2021}. 
    The additional column indicates if the feature is considered in the scrambling process.
    The abbreviation PF stands for Particle Flow, and is related to the event reconstruction algorithm used to process the raw collision data. 
    The output of the algorithm are the so-called PF candidates~\cite{Knapp2021}.
    }
    \label{tab:features}
\end{table}

\section{Scrambling}
\label{app:scrambling}

With the scrambling process we randomly generate data samples that do not conform with the SM, without relying on any BSM theory.
The scrambling is done by replacing the feature values in the original dataset with a new value chosen according to a Gaussian distribution $\mathcal{N}(\mu, \sigma)$, where $\mu = f$ is the initial value of the feature and $\sigma$ is the standard deviation. 
Depending on the feature we apply different strategies on how to choose the standard deviation, and also different post-processing methods to comply with conservation laws and detector limitations.
Therefore, the features are divided into four categories: momenta, isolations, jets and particle numbers, each with their own scrambling strategy.

\subsection{Momenta}
\label{sapp:scrambling_momenta}
There are two different types of transverse momenta we can scramble, the transverse momenta related to the leptons and the transverse momenta related to the jets. 
In theory, due to conservation of momenta, these two types of momenta would be related.
However, for the jets we only have the scalar sum of the transverse momenta as a feature in the dataset ($H_T$). 
We therefore scramble the $H_T$ independently of the transverse momenta of the leptons, and assume that the change in $H_T$ can be absorbed in an appropriate change in the directions of the transverse momenta of the jets.
The $H_T$ is updated according to the following equation,
\begin{equation}
    (H_T)' \sim \left| \mathcal{N}(H_T, \lambda_{H_T} H_T) \right| \,.
\end{equation}
For the transverse momenta related to the leptons we only scramble the transverse momentum $\bm{p}^l_T$ of the chosen lepton $l$ with the highest transverse momentum.
The features describing $\bm{p}^l_T$, are the transverse momentum $p^l_T$ and the pseudorapidity $\eta^l$, where $l \in \{e, \mu\}$ is either an electron or a muon. 
Changing these two features has an effect on other features, which have to be adapted accordingly. 
Specifically, the parallel and orthogonal component of the missing transverse momentum ($\text{MET}_{\parallel}$, $\text{MET}_{\bot}$), and the sum of the transverse momenta of the leptons $p^l_{T,\text{TOT}}$.
The total momentum of the lepton $l$ can be written as
\begin{equation}
    \bm{p}^l = |\bm{p}^l| \begin{pmatrix}
        \sin\theta \cos\phi \\[1pt]
        \sin\theta \sin\phi \\[1pt]
        \cos\theta 
    \end{pmatrix} = \begin{pmatrix}
        p^l_T \cos\phi \\[1pt]
        p^l_T \sin\phi \\[1pt]
        p^l_L
    \end{pmatrix} \,,
\end{equation}
where the coordinate system is chosen such that the $z$-axis is along the beam line, and the transverse momenta lies in the $x$-$y$ plane.
The polar angle $\theta$ is the angle between $\bm{p}$ and the $z$-axis, the azimuthal angle $\phi$ is the angle between $x$-axis and the transverse momentum, $p^l_T$ is the transverse momentum, and $p^l_L$ is the longitudinal momentum. 
The pseudorapidity $\eta^l$ is defined as a function of the polar angle, $\eta^l = -\log\left(\tan{\theta/2}\right)$. 
As an initial point we set $\phi = 0$ as it is possible to align the $x$-axis with the transverse momentum $\bm{p}^l_T$. 
In this case, the momentum $\bm{p}^l$ is fully determined by the transverse momentum $p^l_T$ and the pseudorapidity $\eta^l$,
\begin{equation}
    \bm{p}^l = p^l_T \begin{pmatrix}
        1 \\
        0 \\
        \frac{1}{\tan(\theta(\eta^l))}
    \end{pmatrix} \,,
\end{equation}
where the polar angle $\theta$ is determined by the pseudorapidity $\eta^l$. 
To sample a new momentum vector for lepton $l$ we randomly generate values for the following three quantities,
\begin{equation}
    \begin{aligned}
        (p_T^l)' &\sim \mathcal{N}(p_T^l, \lambda_p p_T^l)\,, \\
        \phi' &\sim \mathcal{N}(0, \lambda_\phi)\,, \\
        (\eta^l)' &\sim \mathcal{N}(\eta^l, \lambda_\eta \sigma_\eta)\,.
    \end{aligned} 
\end{equation}
The sampled transverse momentum $p_T^l$ has to fulfill the requirement $(p_T^l)' > 23$ GeV, and the sampling is therefore repeated until this constraint is respected.

\medskip

Assigning a new value to the momentum of lepton $l$ has an effect on other quantities, which we have to adapt accordingly:
\begin{itemize}
    \item \textit{Missing transverse energy}: The missing transverse energy $\text{MET}$ is specified by a vector with two components
    \begin{equation}
        \begin{pmatrix}
            \text{MET}_{\parallel} \\
            \text{MET}_{\bot}
        \end{pmatrix} = \begin{pmatrix}
            - \sum_q p_{T,\parallel}^q \\
            - \sum_q p_{T,\bot}^q
        \end{pmatrix} \,,
    \end{equation}
    where the parallel and perpendicular direction are with respect to lepton $l$. 
    Sampling a new transverse momentum $\bm{p}^l_T$ for lepton $l$ also changes the definition of the parallel and perpendicular direction. 
    To update the components of the MET, we therefore first add $p^l_T$ to $\text{MET}_{\parallel}$, rotate the $\text{MET}$ by the sampled azimuthal angle $\phi'$ and add $(p^l_T)'$ to the new parallel component, which results in the following equations,
    \begin{equation}
        \begin{aligned}
            (MET_{\parallel})' =& (MET_{\parallel} + p^l_T) \cos\phi' \\
                                & + MET_{\bot} \sin\phi' - (p^l_T)' \,, \\
            (MET_{\bot})' =& - (MET_{\parallel} + p^l_T) \sin\phi' \\
                           & + MET_{\bot} \cos\phi' \,.
        \end{aligned}
    \end{equation}
    
    \item \textit{Sum of transverse momenta of leptons}:
    In the SM dataset, the vector sum of the transverse momenta of all leptons is only characterized by its absolute value. 
    Therefore, we miss the directional information, required for an accurate compensation of the change in $p^l_T$, and we update the sum of transverse momenta in the following way
    \begin{equation}\label{eq:update_lepton_momenta}
        (p^l_{T,\text{TOT}})' = p^l_{T,\text{TOT}} - p^l_T + (p^l_T)' \,.
    \end{equation}
    This does not consider the direction of the momenta, however, usually $\bm{p}^l_T$ is the dominant component of $\bm{p}^l_{T,\text{TOT}}$ and equation~\eqref{eq:update_lepton_momenta} is a good approximation of the actual change in $p^l_{T,\text{TOT}}$.
\end{itemize}

\subsubsection{Isolations}
\label{sssec:isolations}
The isolation $\text{Iso}$ of the leptons, photons and neutral atoms is randomly assigned a new value according to 
\begin{equation}
    \text{Iso}' \sim |\mathcal{N}(\text{Iso}, \lambda_\text{Iso} \, \sigma_\text{Iso}) | \,.
\end{equation}
The absolute value is taken, because the isolation is always positive. 
Additionally, as a requirement of the reconstruction process of an CMS event, the isolation has to be smaller than 0.45. 
Therefore, the sampling is repeated until this constraint is respected.

\subsection{Jets}
\label{sapp:scrambling_}
The total number of jets $N_J$ and the number of jets involving a b-quark $N_B$ are randomly assigned a new value according to
\begin{equation}
    N'_{J,B} \sim \left| \left[ \mathcal{N}(N_{J,B}, \lambda_{J,B} N_{J,B}) \right] \right| \,.
\end{equation}
The number of jets is a non-negative integer, and therefore we round to the nearest integer (denoted by $[\cdot]$) and take the absolute value. 
Additionally, the number of b-jets cannot exceed the total number of jets. 
Therefore, the sampling is repeated until $N'_b \leq N'_J$.

\subsection{Particle Number}
\label{sapp:scrambling_particle_number}
The particle number $N$ for the neutral and charged hadrons, photons, electrons and muons are assigned a new value according to
\begin{equation}
    N' \sim \left| \left[ \mathcal{N}(N, \lambda_N N) \right] \right| \,,
\end{equation}
where the standard deviation is proportional to the original value. 
The particle number is a non-negative integer, and therefore we round to the nearest integer and take the absolute value.

\subsection{Scrambling factors}
\label{sapp:scrambling_factors}

We create three different anomaly datasets each with a different strength of the scrambling. 
The strength of the scrambling is controlled by the scrambling factors introduced in section~\ref{sec:data_set_scrambling}. 
The three scrambling strength are denoted as low, medium and high.
The specific values of the scrambling factors are listed in table~\ref{tab:scrambling_factors}.

\begin{table}
    \centering
    \begin{tabular}{ l | c | c | c }
        Factor & Low & Medium & High \\
        \hline
        $\lambda_p$ & 0.5 & 1.0 & 2.0 \\
        $\lambda_\phi$ & 0.05 & 0.1 & 0.2 \\
        $\lambda_\eta$ & 0.05 & 0.1 & 0.2 \\
        $\lambda_{H_T}$ & 0.5 & 1.0 & 2.0 \\
        $\lambda_\text{Iso}$ & 0.25 & 0.5 & 1.0 \\
        $\lambda_J$ & 0.5 & 1.0 & 2.0 \\
        $\lambda_b$ & 0.5 & 1.0 & 2.0 \\
        $\lambda_{N,\text{charged}}$ & 0.25 & 0.5 & 1.0 \\
        $\lambda_{N,\text{neutral}}$ & 0.25 & 0.5 & 1.0 \\
        $\lambda_{N,\text{photons}}$ & 0.25 & 0.5 & 1.0 \\
        $\lambda_{N,\text{electrons}}$ & 0.5 & 1.0 & 2.0 \\
        $\lambda_{N,\text{muons}}$ & 0.5 & 1.0 & 2.0
    \end{tabular}
    \caption{Factors determining the scrambling strength for low/medium/high scrambling.}
    \label{tab:scrambling_factors}
\end{table}

\section{Quantum feature map}
\label{app:feature_map}

There exist several different strategies to embed classical data into a quantum state, e.g. the basis encoding, the amplitude encoding, or, generally, the encoding via a quantum feature map~\cite{Schuld2018}.

In our work, we apply the embedding introduced in~\cite{Havlicek2019}, which is conjectured to be hard to simulate classically. The formal definition of the embedding unitary is
\begin{equation}\label{eq:feature_map}
    \mathcal{U}_{\Phi(\bm{x})} = U_{\Phi(\bm{x})} H^{\otimes n} U_{\Phi(\bm{x})} H^{\otimes n} \,,
\end{equation}
where $H^{\otimes n}$ is a layer of Hadamard gates and 
\begin{equation}
    U_{\Phi(\bm{x})} = \exp\left( i \prod_{S \subseteq [n]} \phi_S(\bm{x}) \prod_{i \in S} Z_i \right) \,.
\end{equation}
The encoded state is obtained by applying the embedding unitary to the all zeros state $\ket{0}^{\otimes n}$. We choose the following coefficients $\phi_S$ for the data encoding
\begin{equation}
    \begin{aligned}
        \phi_{\{i\}} &= \gamma x_i \,, \\
        \phi_{\{i, j\}} &= \frac{(\gamma x_i) (\gamma x_j)}{\gamma \pi} = \frac{\gamma}{\pi} x_i x_j \,,
    \end{aligned}
\end{equation}
where we also introduced the hyperparameter $\gamma$, which is a constant scaling factor of the input features $\bm{x}$. 
The rescaling of the two-index coefficients $\phi_{\{i,j\}}$ ensures that all encoded rotation angles lie in the interval $[-\gamma\pi, \gamma\pi]$, if we assume that $x_i \in [-\pi, \pi]$. \\
In section~\ref{sapp:feature_map_entanglement} we consider three different options for the index set of the two-qubit coefficients, which leads to different choices of entanglement between the qubits.

\section{Evaluation of training workflow}
\label{app:training_workflow}

\begin{figure}
    \centering
    \includegraphics[width=0.8\columnwidth]{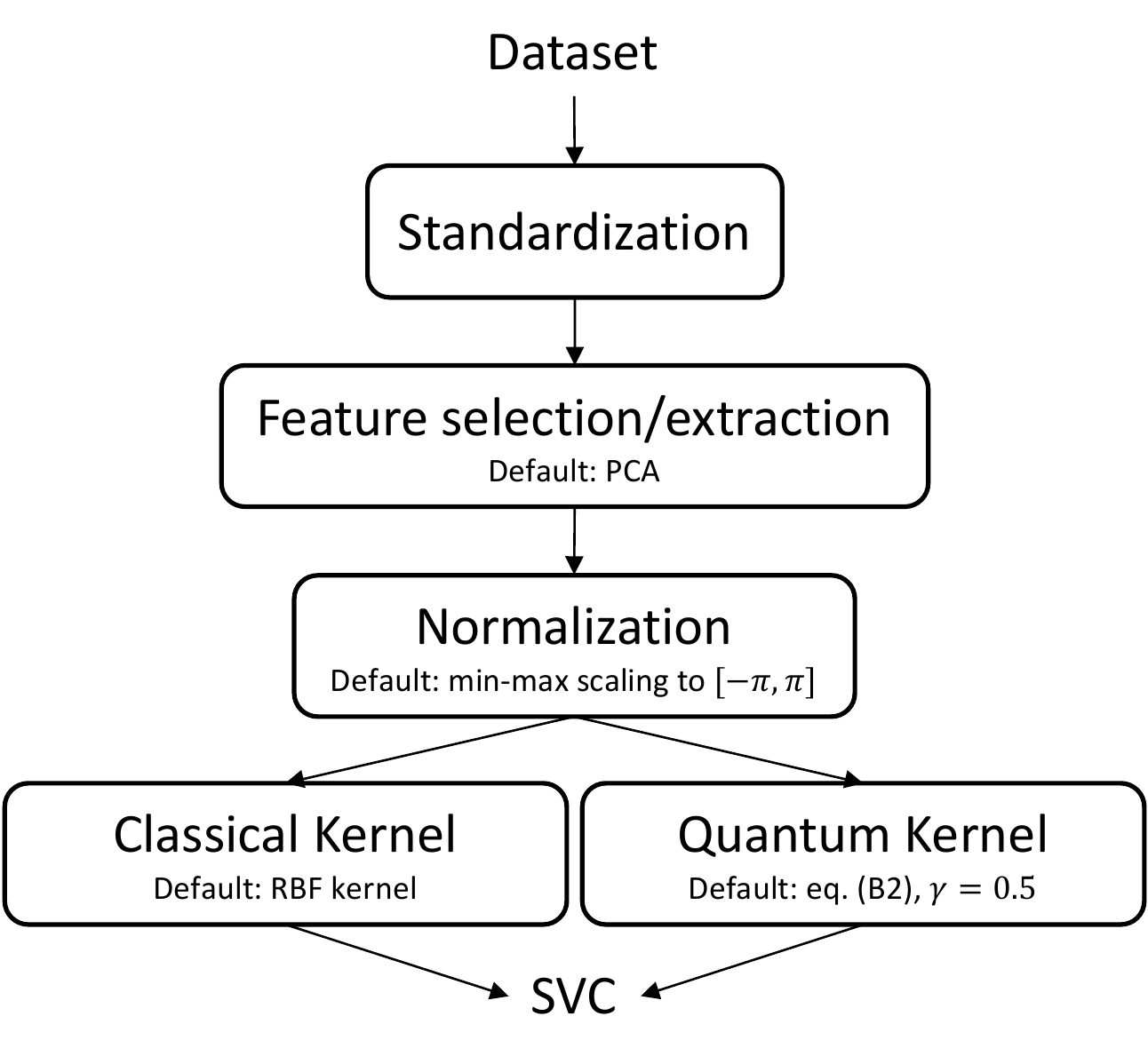}
    \caption{Flowchart of the pipeline used for the training of a quantum or classical SVC.}
    \label{fig:training_workflow}
\end{figure}

In the following we present a detailed evaluation of the different steps in the training workflow of a QSVC. 
The steps are shown as a flowchart in figure~\ref{fig:training_workflow}.
A similar study can also be done for the classical SVC.
However, some steps of the preprocessing of the data for the classical SVC coincide with the steps presented below, therefore we will only focus on the QSVC in the following.

We are going to evaluate the different options at each step of the training workflow. 
To reduce the simulation times, all classifications are done with datasets containing 50 samples per class.
Additionally, we only consider the scrambled dataset with medium scrambling strength.
In all evaluations, a QSVC is trained on 10 different training datasets and validated on one test dataset.

\begin{figure*}
    \centering
    \includegraphics[width=\textwidth]{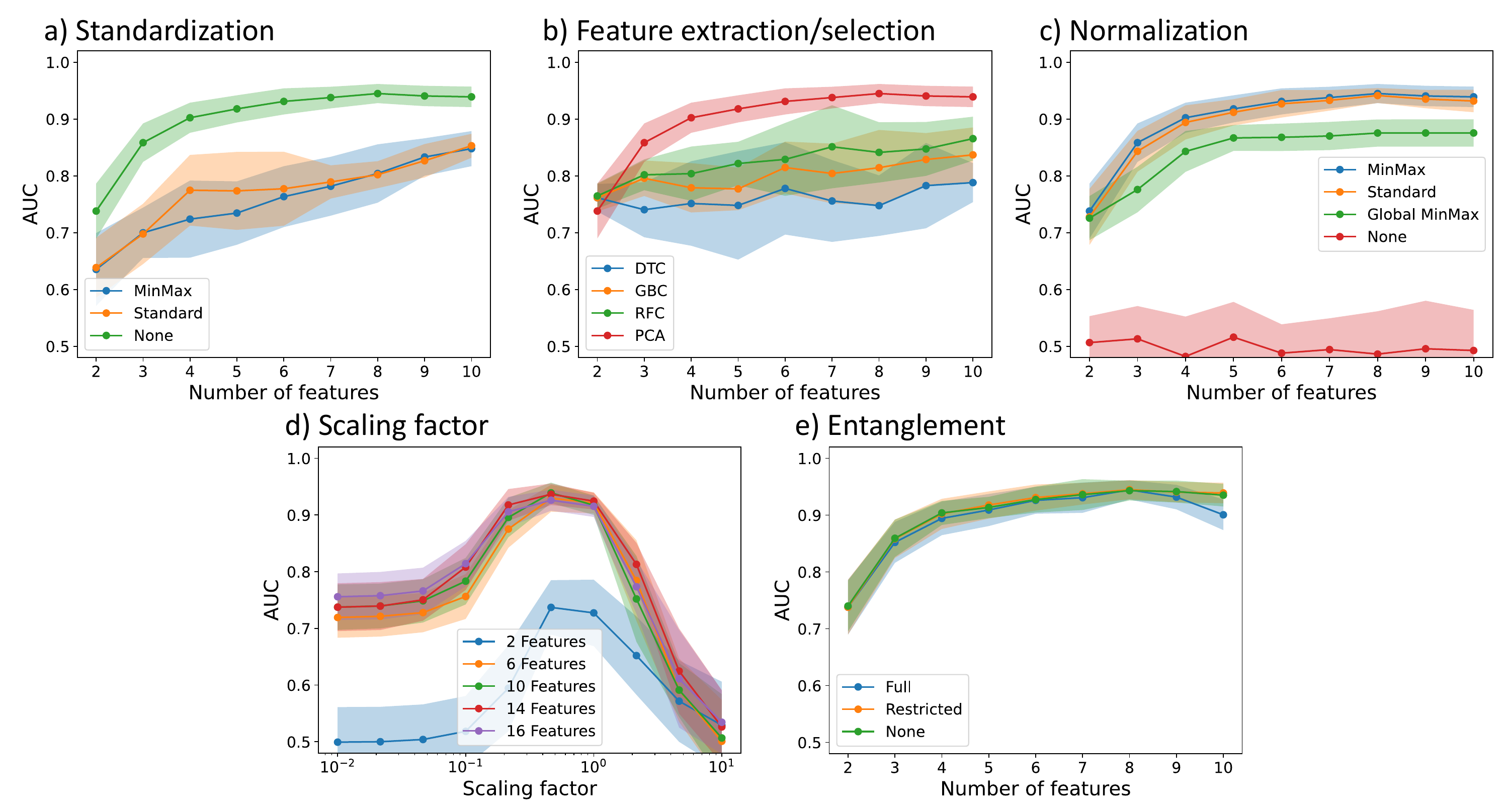}
    \caption{Hyperparameter analysis for the training of a QSVC. Comparing different standardization transformations (\textbf{a}), feature selection and feature extraction methods (\textbf{b}), and normalization transformations (\textbf{c}) for an increasing number of features, searching for the optimal scaling factor (\textbf{d}), and comparing different entanglement strategies of the feature map (\textbf{e}). More detailed information about the specific algorithms and evaluation settings can be found in the text of the corresponding sections.}
    \label{fig:hyperparameter_analysis}
\end{figure*}

\subsection{Standardization}
\label{sapp:standardization}
The standardization is applied to a dataset prior to the feature selection/extraction, and has the purpose to balance the importance of the different features.
We consider two different standardization transformations, the transformation to mean zero and unit variance, and the transformation to the fixed interval $[-1, 1]$, denoted as standard scaling and min-max scaling, respectively. 
As a reference, we also consider the training without any standardization transformation.
For all other steps of the training workflow we choose the defaults given in figure~\ref{fig:training_workflow}.
The results are shown in figure~\ref{fig:hyperparameter_analysis}a.

As expected, the AUC increases with the number of features for all considered standardization options.
Surprisingly, however, we get the best performance when we use no standardization transformation (green curve).
Therefore, we will use no standardization transformation prior to the feature selection/extraction in the following experiments.

\subsection{Feature selection/extraction}
\label{sapp:feature_selection_extraction}
The Principal Component Analysis (PCA) algorithm is a widely used algorithm for feature extraction. 
For comparison, we also consider three feature selection strategies, the decision tree classifier (DTC), the gradient boosting classifier (GBC), and the random forest classifier (RFC).
We use no standardization transformation prior to the feature selection/extraction, and for the remaining steps in the training workflow we choose the defaults given in figure~\ref{fig:training_workflow}.
The results are shown in figure~\ref{fig:hyperparameter_analysis}b.

All feature selection strategies have about the same performance. 
However, all are out-performed by the feature extraction with PCA.
Therefore, we will use PCA to extract the features in the following experiments.

\subsection{Normalization}
\label{sapp:normalization}
The normalization transformation is applied after the feature selection, with a similar purpose as the standardization transformation.
We want to balance the importance of the features before inputting them to the classification algorithm.
We again consider the standard scaling and min-max scaling to the interval $[-\pi, \pi]$.
Additionally, we also consider a global min-max scaling, where the features are transformed to a fixed interval with a joint transformation, instead of individual transformations for each feature. 
The intuition behind the global min-max scaling is that the features retain their relative structure obtained through the feature selection/extraction algorithm.
As a reference, we again consider the training without any normalization transformation.
We use no standardization transformation, and PCA for the feature extraction. 
For the remaining steps of the training workflow we take the defaults given in figure~\ref{fig:training_workflow}.
The results are shown in figure~\ref{fig:hyperparameter_analysis}c.

Using a normalization transformation is clearly beneficial. 
Without the normalization the trained classifier has the same performance as a random classifier (AUC is 0.5).
All other normalization transformations have a similar performance, with the standard and min-max scaling having the highest AUC values.
In the following we will use the min-max scaling to the interval $[-\pi, \pi]$, in order to have some control over the range of values the features will take after the normalization.

\subsection{Scaling factor}
\label{sapp:scaling_factor}
The scaling factor, as introduced in~\cite{shaydulin_importance_2021,peters_machine_2021}, is interpreted as a hyperparameter of the applied quantum kernel. 
This parameter should therefore be optimized.
Here, we show an example of such an optimization, and how the optimal scaling factor is chosen.
For the preprocessing we apply the optimal steps found in the previous sections, and for the feature map we use the default given in figure~\ref{fig:training_workflow}.
The results are shown in figure~\ref{fig:hyperparameter_analysis}d.

The figure shows a sweep over the scaling factor, for different number of features used for the classification. 
Clearly, there is some region of scaling factor values, where the resulting classifier has the best performance.
Additionally, this region is similar for different number of features. 
Of the considered scaling factor values, $\gamma = 0.46$ results in the highest performance for all considered number of features.
For this learning task and this specific choice of feature map a scaling factor of $\gamma \approx 0.5$ therefore leads to a good performance. 
This values of the scaling factor ensures that all angles entering the feature map will effectively be in the interval $[-\frac{\pi}{2}, \frac{\pi}{2}]$. \\
For a different dataset and/or a different feature map the optimal scaling factor may be different.

\subsection{Feature map}
\label{sapp:feature_map_entanglement}
We consider three different options for the entanglement layout of the feature map introduced in section~\ref{app:feature_map}. 
We have \textit{full} entanglement if we apply a two-qubit gates for each pair of qubits. 
We also consider \textit{linear} entanglement is we apply a ``ladder'' of two-qubit gates only for neighbouring qubits (without periodic boundary).
For the $R_{ZZ}$ rotation applied in the feature map, the linear entanglement is equivalent to a more depth efficient layout, where at most two layers of $R_{ZZ}$ are required (even and odd connections between neighbouring qubits).
We call this layout \textit{restricted} entanglement, and apply it here instead of the linear entanglement, which is favourable when running the circuits on hardware.
The last type of layout we consider is the \textit{separable} encoding, without any entanglement between the qubits.
For the preprocessing of the input data, we use the optimized steps presented in the previous sections.
The results are shown in figure~\ref{fig:hyperparameter_analysis}e. 

The figure shows the validation AUC for an increasing number of features for different entangling strategies. 
Against our expectations, the entangling strategy does not have a significant influence on the AUC of the classification.
Generally, adding entanglement between the qubits, is expected to reveal correlations among the features, and we therefore would expect an improvement in the AUC for increasing entanglement.
Possible explanations why this can not be observe here, could be that the classification task is too easy, or during the preprocessing step (especially the PCA) all correlations between the extracted features are removed, and adding correlations in form of entanglement is therefore not improving the classification. 
It could also be, that the dataset does not fall into the class of problems where using a quantum model could lead to an advantage over classical models.
We present a corresponding investigation in section~\ref{sapp:potential_qa}.

Although we do not see a practical advantage of using entanglement for the specific datasets used in the evaluation, we still use the restricted entanglement for the experiments in the main text in order to keep the anomaly detection scheme as general as possible, also across different datasets.

\section{Restricting classical model}
\label{sapp:classical_bias}
\begin{figure*}
    \centering
    \includegraphics[width=\textwidth]{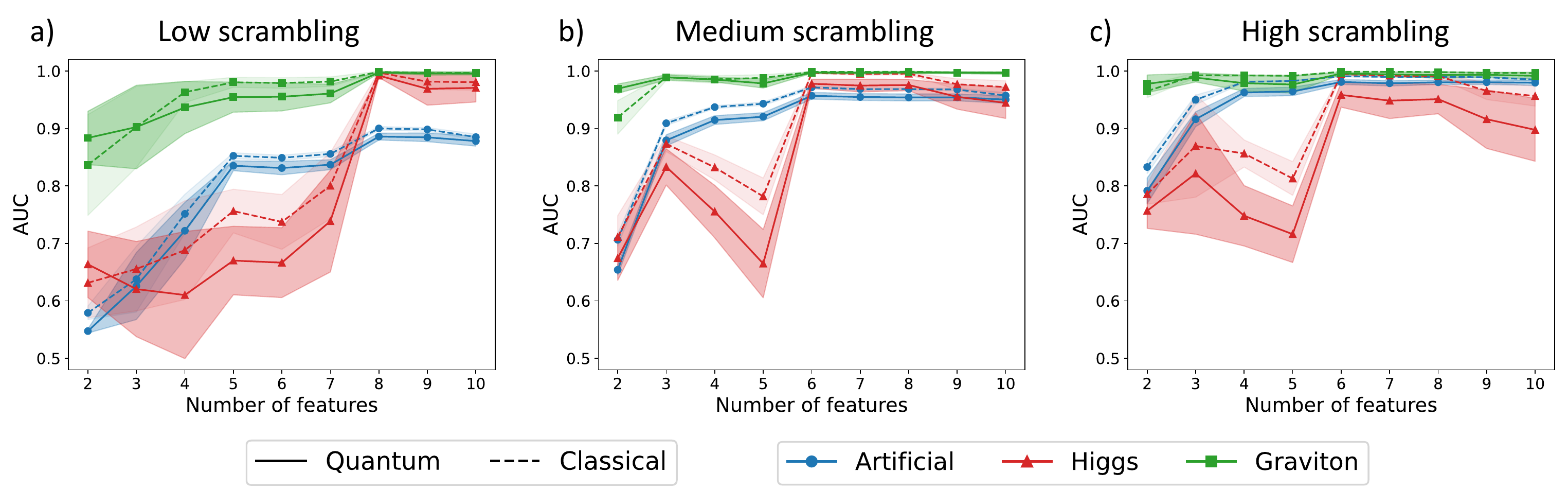}
    \caption{Validation AUC for classification of BSM events among SM events with quantum and classical SVCs for different scrambling strength and different number of features. Same experimental setup as in figure~\ref{fig:results}, except that the hyperparameter of the classical RBF kernel (equation~\eqref{eq:classical_kernel}) is fixed to $\gamma = 0.5$. Note: only the curves of the classical SVC changed (dashed lines) compared to figure~\ref{fig:results}.}
    \label{fig:results_biased}
\end{figure*}

In the main text the hyperparameter of the classical RBF kernel (equation~\eqref{eq:classical_kernel}) was optimized for each combination of training and test dataset.
Here, the hyperparameter will be fixed to $\gamma = 0.5$. 
The resulting classification and detection score are shown in figure~\ref{fig:results_biased}.
Compared to the results in figure~\ref{fig:results} the drop in the classification AUC is minimal, but the detection AUC is significantly improved.
The classical SVC now also outperforms the quantum SVC in the detection.
However, this improvement is expected to be very specific to the classification and detection problem at hand, and cannot be expected in general (especially when the real anomalies are unknown).

\section{``Power of data'' metrics}
\label{sapp:potential_qa}
In Ref.~\onlinecite{Huang2021} the authors introduce a strategy to check if a dataset falls into the class of problems where quantum ML models may perform better than classical models.
In the following, we will follow this strategy to check the class of the studied HEP dataset.
The metrics introduced in Ref.~\onlinecite{Huang2021} were calculated using the code provided in the software package QuASK~\cite{marcantonio2022quask}.
First, we evaluate the geometric difference $g$, defined as a similarity measure between two kernel matrices,
\begin{equation}\label{eq:geometric_difference}
    g_{\text{gen}} = \sqrt{\left\| \sqrt{K_1} \sqrt{K_2} (K_2 + \lambda I)^{-2} \sqrt{K_2} \sqrt{K_1} \right\|_{\infty}} \,,
\end{equation}
where $\lambda$ is a regularization parameter. The geometric difference $g_{\text{gen}}$ is also related to a training error, which is upper bounded by
\begin{equation}\label{eq:training_error}
    g_{\text{tra}} = \lambda \sqrt{\left\| \sqrt{K_1} (K_2 + \lambda I)^{-2} K_1 \right\|_{\infty}} \,.
\end{equation}
Similarly to Ref.~\onlinecite{Huang2021} we report the geometric difference $g_{\text{gen}}$ for a $\lambda$ such that the training error $g_{\text{tra}} \approx 0.0045$. 

\begin{figure*}
    \centering
    \includegraphics[width=\textwidth]{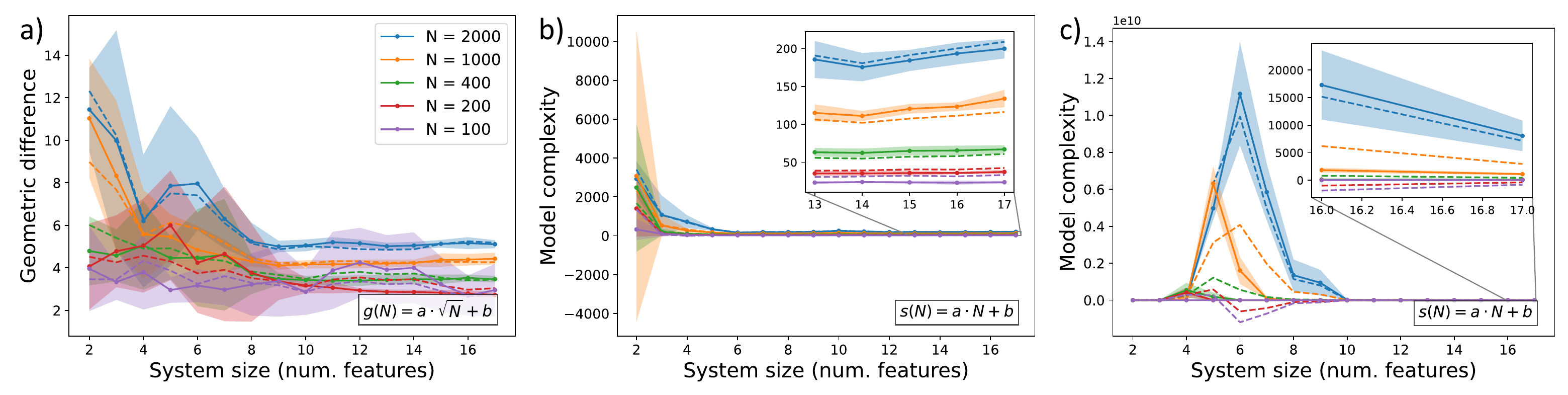}
    \caption{Metrics for the characterization of the ``hardness'' of classifying SM and artificial anomalies. \textbf{a} Geometric difference between classical and quantum kernels for difference dataset sizes and increasing number of features. \textbf{b} Model complexity of classical kernels for different dataset sizes and increasing number of features. The inset shows the model complexities for 13 up 17 features. \textbf{c} Model complexity of quantum kernels for different dataset sizes and increasing number of features. The inset shows the model complexities only for 16 and 17 features.}
    \label{fig:geometric_difference}
\end{figure*}

In figure~\ref{fig:geometric_difference}a we show the geometric difference, calculated from kernel matrices obtained with the workflow described in the previous sections, for the classification between the SM and anomalies (resulting from the medium scrambling) for different numbers of features and different number of samples $N$ in the training and validation datasets. 
For each problem instance (specific number of features and samples), the geometric difference is calculated between the corresponding classical and quantum kernel matrix.
In the figure, we additionally fit a function proportional to $\sqrt{N}$ to the geometric differences (dashed lines), specifically $g(N) = a \cdot \sqrt{N} + b$.
The fit is done separately for each number of features.
Visually, there is good agreement between the measured geometric differences and the dashed lines, which puts us in the regime where the geometric difference scales proportional to $\sqrt{N}$.

After finding the scaling proportional to $\sqrt{N}$, the next step in the assessment is to calculate the model complexity, defined as 
\begin{equation}\label{eq:model_complexity}
    s_K(N) = \sum_{i=1}^N \sum_{j=1}^N \left(\sqrt{K} (K+\lambda I)^{-2} \sqrt{K}\right)_{ij} \, y_i y_j \,,
\end{equation}
where $y_{i,j}$ are the labels of the classification, $K$ is the classical or quantum kernel matrix, and $\lambda$ is again a regularization factor.
Related to the model complexity we can define a training error,
\begin{equation}\label{eq:training_error_model_complexity}
    t_K(N) = \lambda^2 \sum_{i=1}^N \sum_{j=1}^N \left( \left(K + \lambda I \right)^{-2} \right)_{ij} \, y_i y_j \,.
\end{equation}
We choose the regularization such that this training error and the model complexity are both minimized.
The resulting model complexities for the classical kernel matrices and the quantum kernel matrices are shown in figure~\ref{fig:geometric_difference}b and~\ref{fig:geometric_difference}c, respectively.
In both cases, we fit a function proportional to $N$ to the model complexities (dashed lines), specifically $s(N) = a \cdot N + b$. 
Visually, we observe good agreement between the measured model complexities and the dashed lines especially for the higher number of samples, at least for the classical SVC.
For the quantum SVC, the calculation of the model complexity is somehow not very meaningful.
The model complexity is orders of magnitude bigger than in the classical case, and the values also seem to converge for larger numbers of features.
However, they do not seem to be fully converged yet, and a fair assessment is not possible.
Therefore, based on the geometric difference and the classical model complexity, we could end up either in the case with ``Potential quantum advantage'' or the case where the problem is ``Likely hard to learn''.

From the results we obtained in the sections above, we would have expected to be in the case where both, classical and quantum SVCs, can learn well (either $g_{CQ} \ll \sqrt{N}$ or $s_C \ll N$).
Therefore, the results obtained from the measured metrics is not conclusive and this test cannot be used to argue for (or against) quantum advantage in the studied classification problem.

\section{Hardware experiments}
\label{app:ibmq_hardware}

\begin{figure}
    \centering
    \includegraphics[width=0.8\columnwidth]{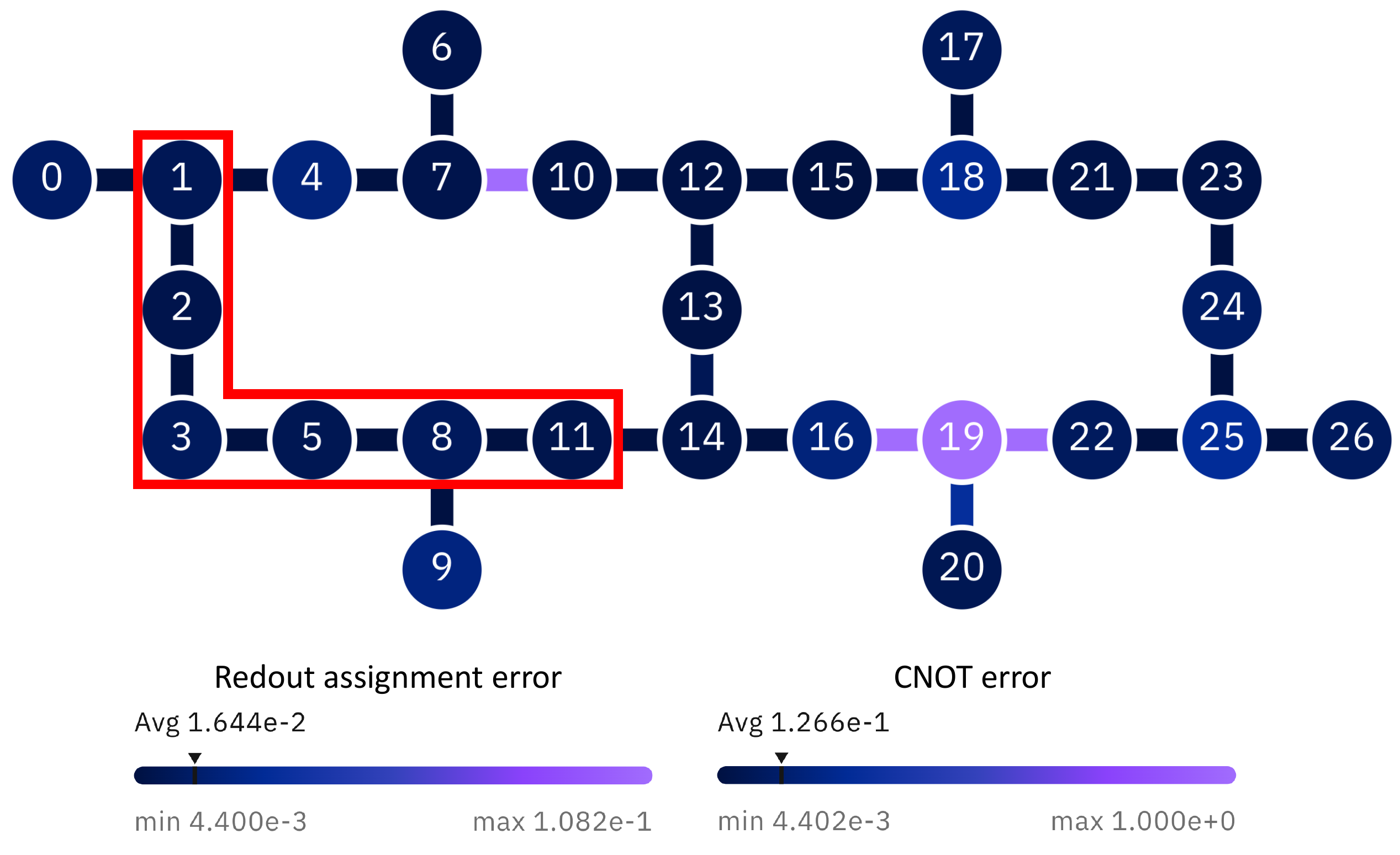}
    \caption{Qubit Layout of \textit{ibm\_auckland}. 
    The highlighted qubits are the ones used in the experiments. 
    The color of the qubits denotes the readout assignment error, and the color of the edges the CNOT error (colors obtained on October 04, 2022). 
    The highlighted 6 qubits are the ones used to estimate the kernel values.}
    \label{fig:ibm_auckland}
\end{figure}

For the hardware experiments we used the IBM Quantum device \textit{ibm\_cairo}. 
The qubit layout of the device is shown in figure~\ref{fig:ibm_auckland}. 
To mitigate some of the hardware errors we apply a depolarization error mitigation strategy to the obtained kernel matrices.

\subsection{Depolarization error mitigation}
\label{sapp:depol_mitigation}
Some of the errors occurring on the hardware can be modeled by a depolarization channel
\begin{equation}\label{eq:depol_channel}
    \mathcal{D}_\lambda [\rho] = \lambda \rho + (1 - \lambda) \frac{1}{2^n} \,,
\end{equation}
where $\lambda$ is the survival probability of the original quantum state $\rho$, and $n$ is the number of qubits. 
To mitigate the depolarization error we can exploit that in the noiseless kernel matrix $K$ all diagonal entries are 1.
Therefore, if one measures the diagonal entries in a noisy settings one can gather information about the device noise~\cite{Hubregtsen2021training}.
The survival probability $\lambda_i$ of the noisy kernel matrix element $K^*_{ii}$ is
\begin{equation}
    \lambda_i = \sqrt{\frac{K_{ii}^* - 2^{-n}}{1 - 2^{-n}}} \,.
\end{equation}
The mitigated kernel values can then be obtained with
\begin{equation}
    K_{ij} = \frac{K_{ij}^* - 2^{-n}(1 - \lambda_i \lambda_j)}{\lambda_i \lambda_j} \,.
\end{equation}
For the experiment in the main text, we assume that all survival probabilities $\lambda_i$ have the same value, which can be estimated with
\begin{equation}
    \lambda = \frac{1}{N} \sum\limits_{i=1}^N \lambda_i \,,
\end{equation}
where $N$ is the size of the symmetric training kernel matrix.
This value can then also be used for the mitigation of the non-symmetric validation kernel matrix.

\end{document}